\documentclass{mem}

\usepackage{natbib}
\usepackage{txfonts}
\usepackage{balance}
\usepackage{graphicx}
\usepackage{graphics}
\usepackage{epsf,epsfig}
\usepackage[dvips]{color}

\usepackage{rotate}
\usepackage{color}
\usepackage{amssymb,amsfonts}
\usepackage{latexsym}

\usepackage[a4paper]{hyperref}

\idline{75}{282}

\begin{document}


\newcommand{\un}[1]{~\hspace{-1pt}\ensuremath{\mathrm{#1}}}
\newcommand{\ti}[1]{$^{44}{}${#1}}
\newcommand{\ca}[1]{$^{40}{}${#1}}
\newcommand{\nic}[1]{$^{56}{}${#1}}
\newcommand{\co}[1]{$^{57}{}${#1}}
\newcommand{\fe}[1]{$^{60}{}${#1}}


\newcommand{\simbolx}{{\it SIMBOL-X}}

\newcommand{\integ}{{\it INTEGRAL}}
\newcommand{\ibis}{IBIS}
\newcommand{\spi}{SPI}
\newcommand{\isgri}{ISGRI}
\newcommand{\ibisgri}{IBIS/ISGRI}

\newcommand{\rxte}{{\it RXTE}}
\newcommand{\xmm}{{\it XMM-Newton}}
\newcommand{\chandra}{{\it Chandra}}
\newcommand{\sax}{{\it BeppoSAX}}
\newcommand{\asca}{{\it ASCA}}

\newcommand{\gro}{{\it CGRO}}
\newcommand{\comptel}{COMPTEL}

\newcommand{\swift}{{\it SWIFT}}
\newcommand{\bat}{BAT}
\newcommand{\vla}{VLA}


\newcommand{\am}{$^{\prime}$}
\newcommand{\as}{$^{\prime\prime}$}
\newcommand{\gammaray}{$\gamma$-ray}
\newcommand{\gammarays}{$\gamma$-rays}
\newcommand{\xray}{X-ray}
\newcommand{\xrays}{X-rays}


\def\ks{km s$^{-1}$}
\def\kms{$\mathrm {km s}^{-1}$}
\def\d{$^\circ$}
\def\m{$^\prime$}
\def\s{$^{\prime\prime}$}
\def\hh{$^{\mathrm h}$}
\def\mm{$^{\mathrm m}$}
\def\ss{$^{\mathrm s}$}
\def\cm3{cm$^{-3}$}
\def\pp{$^{\prime\prime}$}
\def\msun{M$_\odot$}
\def\ha{H$\alpha$}

\def\eg{{\it e.g.~}}
\def\etal{et~al.~}
\def\ie{{\em i.e.~}}

\title{
\ti{Ti} nucleosynthesis gamma-ray lines with \simbolx
}

\subtitle{}

\author{
M. \, Renaud\inst{1}
\and F. \, Lebrun\inst{2,3}
\and A. \, Decourchelle\inst{2,4}
\and R. \, Terrier\inst{3}
\and J. \, Ballet\inst{2,4}
          }

  \offprints{M. Renaud}

\institute{ Max-Planck-Institut f\"ur Kernphysik, P.O. Box 103980,
D69029 Heidelberg, Germany \\
\email{mrenaud@mpi-hd.mpg.de}
\and
Service d'Astrophysique, DAPNIA/DSM/CEA, 91191 Gif-sur-Yvette,
France
\and
APC-UMR 7164, 11 Place M. Berthelot, 75231 Paris,
France
\and
AIM-UMR 7158, 91191 Gif-sur-Yvette, France }

\authorrunning{Renaud}

\titlerunning{\ti{Ti} lines with \simbolx}

\abstract{In this contribution we discuss the \ti{Ti}
nucleosynthesis \gammaray~lines and their visibility with
\simbolx~from simulations based on its expected sensitivity and
spectro-imaging capabilities. The \ti{Ti} radioactive nucleus can
provide invaluable information on the details of supernova
explosions. Its lifetime of $\sim$ 85 yrs makes it the best
indicator of the youth of these stellar explosions through its
three \gammaray~lines at 67.9, 78.4\un{keV} and 1.157\un{MeV}. We
focus on the youngest Galactic supernova remnants, namely:
Cassiopeia~A, for which the location and Doppler-velocity
estimates of the \ti{Ti}-emitting regions in the remnant would
offer for the first time a unique view of nucleosynthesis
processes which occurred in the innermost layers of the supernova;
SN~1987A, in the Large Magellanic Cloud, whose progenitor is known,
and for which the expected measurement of these lines would greatly
constrain the stellar evolution models; Tycho and Kepler SNRs for
which \ti{Ti} lines have never been detected so far. The issue of
the "young, missing and hidden" supernova remnants in the Galaxy
will also be addressed using \simbolx~observations at the position
of the \ti{Ti} excesses that wide-field instruments like those
onboard \integ~and \swift/\bat~should be able to reveal.

\keywords{gamma rays: observations -- ISM: individual
(Cassiopeia~A, Tycho, Kepler, SN~1987A) -- nuclear reactions,
nucleosynthesis, abundances -- supernova remnants } }

\maketitle{}

\section{Introduction}

Gamma-ray lines (from several tens of\un{keV} to\un{MeV} energies)
emerging from radioactive nuclei are the unique way to provide
isotopic information from sources and sites of cosmic
nucleosynthesis and to probe several underlying key physical
processes \citep[see \eg][for a recent review]{c:diehl06bis}.
Moreover, their highly penetrating nature allows astronomers to
probe regions invisible at other wavelengths. The main candidate
sources are novae, supernovae (hereafter, SNe), winds from massive
stars and AGB (Asymptotic Giant Branch) stars. Unfortunately, only
a few of these radioactive nuclei are accessible to
\gammaray~astronomy. The main reason is that the emerging
\gammaray~lines from short-lived isotopes, however
important their activity, might still be blocked in the stellar
interiors and then inaccessible to observation, whereas the
long-lived isotopes must be abundantly produced for their
respective lines to be detectable. Therefore \gammaray~telescopes
which have been exploring this astronomical window for the last
three decades usually deal with very low signal-to-noise ratio
levels.

Most of these observable radioactive isotopes (\eg \nic{Ni},
\co{Co}, \fe{Fe}) are synthesised in SNe, either in core-collapse
or thermonuclear explosions. Amongst them, \ti{Ti} is thought
to be exclusively produced in those events. It emits three
\gammaray~lines with similar branching ratios at 67.9 and
78.4\un{keV} (from \ti{Sc}$^{\star}$) and at 1.157\un{MeV} (from
\ti{Ca}$^{\star}$) during the decay \ti{Ti}
$\longrightarrow$ \ti{Sc} $\longrightarrow$ \ti{Ca} with a
weighted-average lifetime of 85.3 $\pm$ 0.4 yrs \citep[see][and
references therein, for the recent \ti{Ti} lifetime
measurements]{c:ahmad06}. Even though the production site of this
element during the first stages of the explosion is still debated
\citep{c:young06}, the region where the $\alpha$-rich freeze-out
from Si-burning at nuclear statistical equilibrium efficiently occurs
is widely considered as the main location of \ti{Ti} production
since the work of \cite{c:woosley73}. Therefore, this isotope
probes deep into the interior of these exploded stars, provides a
direct way to study the SN-explosion mechanism itself, and
represents the unique signature of any previously unknown young
(\ie few centuries old) supernova remnant (hereafter, SNR) as
well.

As a consequence, \ti{Ti} production is strongly dependent on the
explosion details, specifically on the mass-cut in core-collapse
SNe (the mass above which matter is ejected), which is artificially
defined in most of the theoretical models \citep{c:ww95}, the
energy of the explosion, and the asymmetries \citep[see
\eg][]{c:maeda06} which have been revealed in several SNRs by
recent optical and \xray~observations
\citep{c:laming03,c:fesen06bbis}. Moreover, the production of
elements during the explosion is computed through a large nuclear
network and therefore highly dependent on the nuclear cross sections
of the implied reactions. It is of interest to point out that some of
these nuclear reactions which govern the nucleosynthesis of \ti{Ti}
\citep{c:the98} have been seriously revisited only recently
\citep{c:sonzogni00,c:horoi02,c:nassar06}. It is therefore not
surprising to see a large variation in the predicted theoretical
\ti{Ti} yields, ranging from zero to a few 10$^{-4}$ \msun~for the
most frequent type II \citep{c:ww95,c:tnh96,c:rhhw02,c:lc03} and
type I$_{b/c}$ \citep{c:wlw95} SNe, and up to 10$^{-3}$-10$^{-2}$
\msun~for the rare event of the He-detonation of a
sub-Chandrasekhar white dwarf \citep{c:woosley94,c:livne95}. As
reported by \cite{c:iwamoto99}, \ti{Ti} yields for standard type
Ia SNe would lie between a few 10$^{-6}$ and 5 $\times$ 10$^{-5}$
\msun, in agreement with the improved multi-dimensional models of
\cite{c:travaglio04} and \cite{c:ropke05}.

\begin{figure}[ht!]
\resizebox{\hsize}{!}{\includegraphics[clip=true]{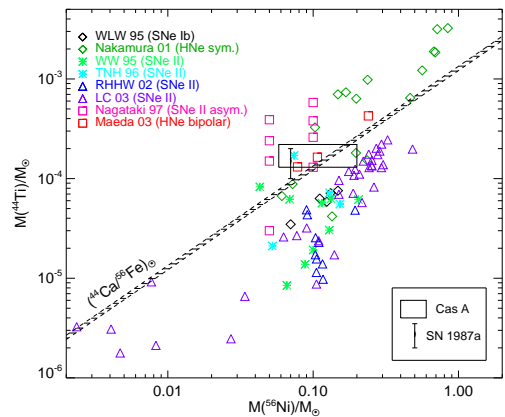}}

\caption{\footnotesize \ti{Ti} yield versus \nic{Ni} yield diagram
from models and observations for the case of core-collapse
explosions. \nic{Ni} is supposed to be responsible for the
early-time optical light-curve and is also thought to be produced
in the innermost layers of the explosion. The black box and line
represent the measurements (or estimates) of yields of these two
nuclei in Cassiopeia~A and SN~1987A, respectively. The different
symbols refer to the main nucleosynthesis models. Only those which
include asymmetries during the explosion would explain the high
\ti{Ti}/\nic{Ni} ratios observed in these SNRs. The dashed area
corresponds to the solar ratio of the corresponding stable
isotopes (\ti{Ca}/\nic{Fe})$_{\odot}$
\citep{c:anders89,c:lodders03}.}

\label{f:fig1}

\end{figure}

Given these large uncertainties in the theoretical estimates,
\gammaray~observations turn out to be of great importance. The
discovery of the 1157\un{keV} \ti{Ca} \gammaray~line emission in
the youngest known Galactic SNR Cassiopeia~A (hereafter, Cas~A)
with \gro/\comptel~\citep{c:iyudin94} was the first direct proof
that this isotope is indeed produced in SNe. This has been
strengthened by the \sax/PDS detection of the two blended low
energy \ti{Sc} lines at 67.9\un{keV} and 78.4\un{keV}
\citep{c:vink01}. Recently, \cite{c:renaud06e} reported the
detection of these two \ti{Sc} lines with the
\integ~\ibisgri~instrument and calculated a weighted-average \ti{Ti}
yield of 1.6 $^{+0.6}_{-0.3}$ $\times$ 10$^{-4}$ M$_{\odot}$. This
high value, as compared to the theoretical predictions, could be due
to several effects such as asymmetries \citep{c:vink04,c:hwang04},
as shown in Figure \ref{f:fig1}, and/or a high energy
of the explosion \citep{c:laming03}. Spectro-imaging measurements
of these \gammaray~lines would then bring unique information about
the location and the dynamics of the \ti{Ti} emission regions.
Moreover, Cas~A appears to be the only SNR from which \ti{Ti} line
emission has been unambiguously
detected\footnote{RX~J0852-4622/GRO~J0852-4642 (alias Vela Junior)
might be the second SNR from which \ti{Ti} \gammaray~line has been
measured. Unlike Cas~A, the detection originally claimed by
\cite{c:iyudin98} with \gro/\comptel~is however still
controversial \citep{c:schonfelder00}.}. Therefore, a significant
gain in sensitivity in the hard \xray/soft \gammaray~domain is
also needed to address other sources of interest (listed by order
of importance): SN~1987A, whose progenitor is known, and for which
the same amount of \ti{Ti} as in Cas~A is expected; Tycho and
Kepler SNRs, most likely the remnants of type Ia SNe, from which
no \ti{Ti} line emission has been detected so far; and the other
\ti{Ti}-source candidates (\ie young SNRs) that
\integ~\ibisgri~and \spi~and \swift/\bat~will list at the end of
their Galactic Plane surveys.

We present here the first simulations of the \ti{Sc}
\gammaray~lines at 67.9 and 78.4\un{keV}, based on the
spectro-imaging capabilities of \simbolx~(see the contributions of
Ferrando \etal 2007, Pareshi \etal 2007 \& Laurent \etal 2007 in
these proceedings). \simbolx~will be the first formation flight
mission whose payload is made of an optics module on a first
spacecraft which focuses the radiation in the 0.5-100\un{keV}
energy range on two detectors nominally placed 20\un{m} away on a
second satellite. This mission will offer for the first time
"X-ray"-like angular resolution ($\sim$ 18\s~HEW) and sensitivity
(more than one order of magnitude better than \ibisgri~up to $\sim$
70\un{keV}) in the hard \xray~domain.

\section{Simulations}

In this contribution, we focus on the lower detection layer, a
Cd(Zn)Te detector, operating between 5 and 100\un{keV}, with a
spectral resolution of the order of 1\un{keV} (FWHM) at the
\ti{Sc} line energies. All the simulations presented here were
carried out with the generic set of response files provided by the
organisers of the workshop.

\subsection{Cas~A: where is the \ti{Ti} ?}

\begin{figure*}[!htb]
\resizebox{\hsize}{!}{\includegraphics[clip=true]{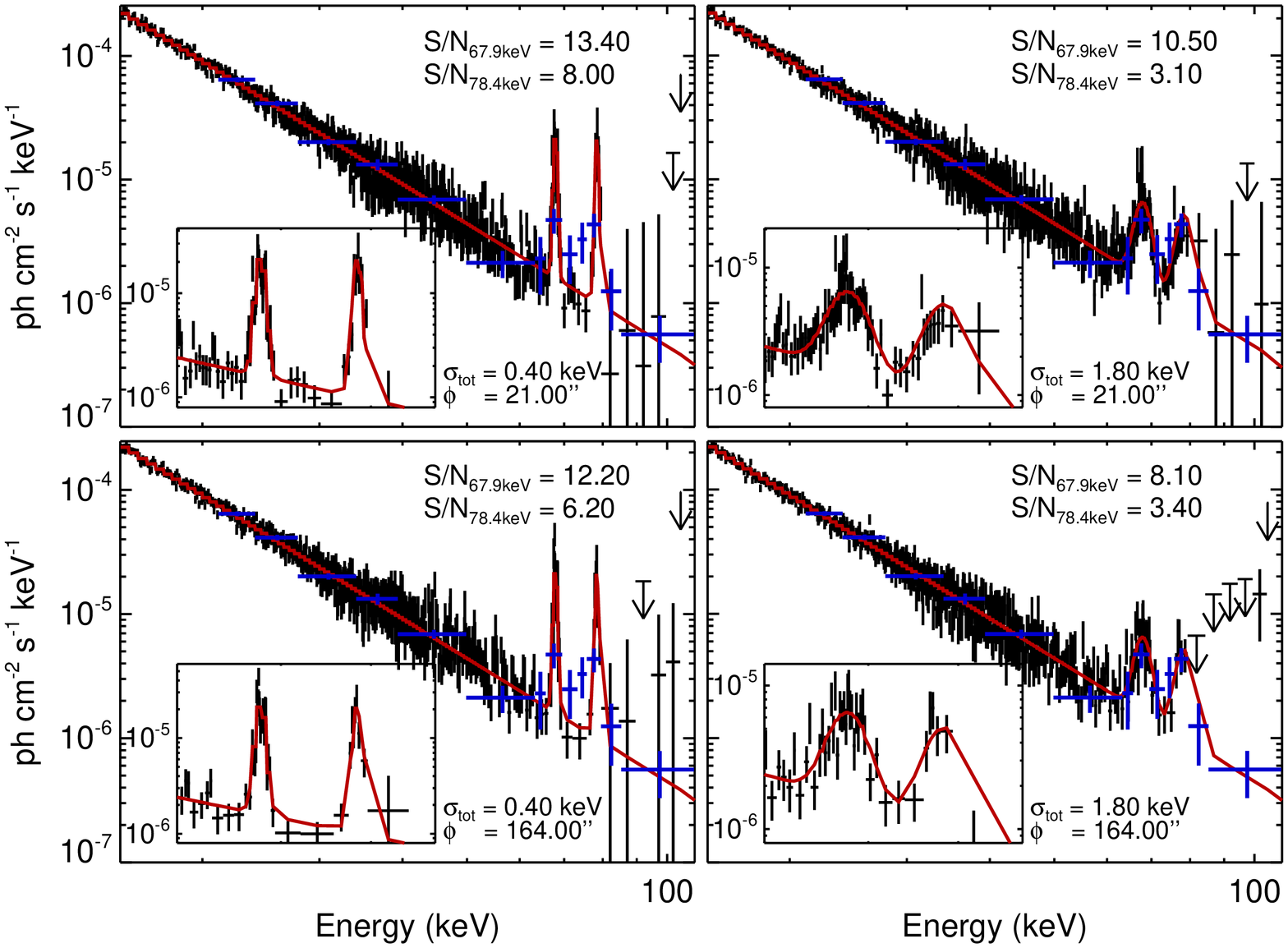}}

\caption{\footnotesize Simulated spectra of Cas~A based on the
\ibisgri~results for an exposure time of 100\un{ks} with
\simbolx~in four scenarios (in black). The case of an extended
source ($\Phi$ = 164\s) correponds to an average velocity of 4000
\ks~over 330\un{yrs} at 3.4\un{kpc}. The case of a broad
line ($\sigma$ = 1.8\un{keV}) refers to an intrinsic velocity of
$\sim$ 8500 \ks, as measured by \cite{c:fesen06abis} in
some fast-moving optical knots. Note that all of the signal-to-noise
ratios were calculated over an interval of 1\un{keV} centered on
the lines. Upper limits are given at the 3 $\sigma$ confidence
level. For comparison, the \ibisgri~data points from
\cite{c:renaud06e} are superimposed in blue.}

\label{f:fig2}

\end{figure*}

Cas~A is the youngest known SNR in the Galaxy, located at a
distance of 3.4$^{+0.3}_{-0.1}$\un{kpc} \citep{c:reed95}. The
estimate of the supernova is A.D.~1671.3$\pm$0.9, based on the
proper motion of several ejecta knots \citep{c:thorstensen01}.
However, an event observed by Flamsteed (A.D.~1680) might be
the origin of the Cas~A remnant
\citep{c:ashworth80,c:stephenson02}. Even if it is generally
accepted that Cas~A was formed by the explosion of a massive
progenitor \citep[see \eg][]{c:vink04}, there is still some
debate on the detailed stellar evolution scenario
\citep{c:young06}. The large collection of data from observations
in the radio, infra-red, optical,
\xray~\citep[see \eg][]{c:hwang04} allows us to study its
morphology, composition, cosmic-ray acceleration efficiency and
secular evolution in detail. Most of this information has been
inferred from recent \xray~observations which revealed a complex
morphology of this SNR \citep{c:laming03}, in particular a
jet-counterjet structure in the Si-rich ejecta and several
extremely Fe-rich regions \citep{c:hwang03,c:hwang04}. Highly
Fe-rich ejecta are sites of complete Si burning in the explosion
and may also harbor some of the \ti{Ti} known to have been
synthesised in $\alpha$-rich freezeout. Unfortunately, as pointed
out by \cite{c:hwang03}, only "a few percent of the total Fe
ejected by the explosion is currently visible in X-rays".
Therefore, direct spatially-resolved spectroscopic measurements of
the nucleosynthesis products through the observation of
radioactive \gammaray~lines would unambiguously reveal the
composition of heavy nuclei and their dynamics.

In order to estimate the visibility of the \ti{Sc} \gammaray~lines
with {\it SIMBOL-X}, we assume a \ti{Ti} line flux of 2.5 $\times$
10$^{-5}$ cm$^{-2}$ s$^{-1}$ and a hard \xray~spectrum between 15
and 100\un{keV} described by a power-law with an index $\Gamma$ =
-3.3, as recently measured by \ibisgri~\citep{c:renaud06e}.
Details about the nature of this continuum emission, which is related
to the acceleration processes at the SNR forward shock, and the
expectations with \simbolx~can be found elsewhere in these
proceedings (Decourchelle \etal 2007). We further assume different
scenarios. The most optimistic case is that of a static point-like
source. The opposite corresponds to an extended source of 164\s~in
diameter (\ie for an average velocity of 4000 \ks~over
330\un{yrs} at 3.4\un{kpc}) with a current velocity of 8500 \ks~(\ie
a line width of $\sim$ 1.8\un{keV} at 70\un{keV}), as observed in some
optical knots \citep{c:fesen06abis}. Figure \ref{f:fig2} shows the
simulated Cas~A spectra for a 100\un{ks} observation with
\simbolx~(in black), in comparison with the \ibisgri~data points
(in blue). Even in the worst case of an high-velocity extended
source, both \ti{Sc} lines would be clearly detected by
\simbolx~for this reasonable exposure time. Moreover, from the
spectral fit, we have estimated a 3 $\sigma$ Doppler-velocity
resolution of the order of (2000-2800) \ks~at 67.9\un{keV} and
(4000-4500) \ks~at 78.4\un{keV}, \ie well below the current
forward shock velocity of $\sim$ 5200 \ks~\citep{c:vink98}. This
would allow us to obtain for the first time unique information on
the dynamics of the nucleosynthesis sites which occurred close to
the mass-cut during the first stages of the explosion.

\begin{figure}[ht!]
\resizebox{\hsize}{!}{\includegraphics[clip=true]{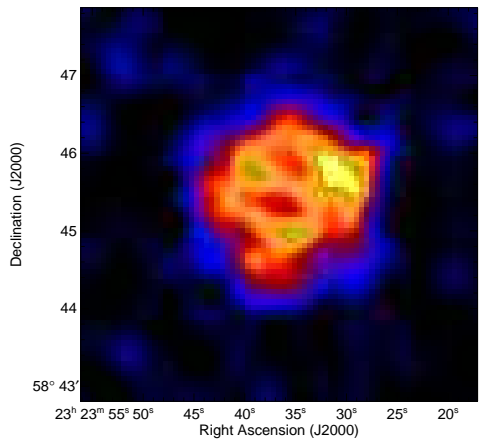}}

\caption{\footnotesize Simulated image of Cas~A with
\simbolx~assuming a uniform distribution of \ti{Ti} within a
82\s~radius sphere and with a global weighted-average line flux
measured by \cite{c:renaud06e}. To improve the visibility of this
low statistics image, a 20\s~Gaussian smoothing has been applied.
At this scale, the \simbolx~PSF (HEW) is five sky pixels.}

\label{f:fig3}

\end{figure}

We have also estimated the imaging capabilities of \simbolx~by
simulating a uniform distribution of \ti{Ti} (note that this would
represent the worst case of detectability) over a 82\s~radius
sphere. The resulting image is shown in Figure
\ref{f:fig3}, on which we applied a 20\s~Gaussian smoothing (FWHM).
\simbolx~should then be able to disentangle amongst several
plausible scenarios of asymmetries originated from the explosion
such as a jet-counterjet structure or a clumpy distribution.

\subsection{SN~1987A: how much \ti{Ti} has been synthesised in the explosion of a known massive star ?}

SN~1987A is unique because it is the only case for which the
progenitor star is known. Sanduleak -69\d202 was initially a red
supergiant of $\sim$ 20 \msun~\citep{c:podsiadlowski92} and
produced a dense, slow, wind focused into the equatorial plane of
the system. The star then evolved into a blue supergiant and began
to produce a high-velocity, low-density, isotropic wind. However,
the details of its late-time evolution are still not completely
clear \citep{c:smith07}, and the presence of a triple-ring optical
nebula around SN~1987A could be due to the merger of two massive
stars 20,000 yrs prior to the explosion \citep{c:morris07}. It now
represents a unique laboratory for studying the interaction between
the circumstellar medium of a massive star and the SN ejecta in
the radio and \xray~domains \citep{c:gaensler07}.

The \gammaray~lines originated from the decay of \nic{Co}
\citep{c:matz88,c:tueller90} and \co{Co} \citep{c:kurfess92} which
are supposed to feed the early-time bolometric light-curve have
been detected from SN~1987A. The current light-curve is thought to
be powered by the radioactive decay of \ti{Ti}. A precise
measurement of its \gammaray~lines would therefore be invaluable
for constraining the models of massive star evolution and
explosive nucleosynthesis (through the \ti{Ti}/\nic{Ni} ratio, as
shown in Figure \ref{f:fig1}). From Monte-Carlo simulations
of Compton degradation of \gammaray~photons in the SN ejecta,
\cite{c:mochizuki04} have found a yield of (0.8-2.3) $\times$
10$^{-4}$ \msun~of \ti{Ti} to be required to power the observed
light-curve. This value is in a good agreement with the modeling
of the infrared emission lines of the ejecta \citep{c:fransson02}.
Such a yield, comparable to that measured in Cas~A, would produce
a \ti{Ti} line flux from SN~1987A (4-12) times weaker than that
measured in the former. Therefore, a 10$^{6}$ s
\simbolx~observation should lead to the detection of the 67.9 and
78.4\un{keV} \ti{Sc} lines at $\sim$ 6 and 4 $\sigma$,
respectively in the worst case (smaller yield).

\subsection{Tycho \& Kepler SNRs: \ti{Ti} yields and nature of the explosions}

The Tycho (SN~1572) and Kepler (SN~1604) SNRs are the remnants of
famous historical Galactic SNe. The stellar explosions have been
observed by several astronomers, from China, Korea and Europe.
Nowadays, both remnants are studied in details through
observations in the radio and \xray~domains. However, many
questions still remain unanswered.

Kepler's progenitor is somewhat of an enigma. The light-curve
obtained by Koreans and Johannes Kepler suggests a type Ia event
with an absolute magnitude M$_{V}$ $\sim$ -18.8 at the peak
\citep{c:stephenson02}, and its distance far above the Galactic
Plane, $>$ 500\un{pc} at a distance of 4.8\un{kpc}
\citep{c:reynoso99}, also supports this hypothesis. Moreover,
\cite{c:kinugasa94} have measured from \asca~observations a
relative overabundance of iron that agrees with type Ia
nucleosynthesis models. However, the nitrogen overabundance in the
optical knots \citep{c:dennefeld82}, the slow expansion velocities
of these knots and the enhanced density in the region suggest that
there is circumstellar material ejected by the stellar wind from a
massive star. \cite{c:bandiera87} has then proposed a model
suggesting that the progenitor was a massive "runaway star"
ejected from the Galactic Plane. Recent \chandra~observations have
confirmed the prominence of iron emission, together with the
absence of O-rich ejecta, making the scenario of a type Ia SN
progenitor with a circumstellar interaction more likely
\citep{c:reynolds07}. Tycho is considered as the prototype of a
type Ia SN \citep{c:baade54}. With an age of 435 yr and a distance
of 2.3 $\pm$ 0.8 kpc \citep{c:smith91}, this SNR is the most
promising candidate to observe explosive nucleosynthesis products
from a thermonuclear SN. \cite{c:decourchelle01} and
\cite{c:hwang02} have studied through \xmm~and
\chandra~observations the composition and the structure of the
ejecta. Similarly to Kepler SNR, Si and Fe-rich ejecta knots have
been found and both integrated \xray~spectra look alike, although
Tycho would be the result of a convential type Ia encountering
low-density, more or less uniform ISM.

No \ti{Ti} \gammaray~line has ever been detected in these two SNRs
so far. Only upper limits from Tycho were reported with
\ibisgri~\citep{c:renaud06b}, from which sub-Chandrasekhar models
have been ruled out. Unfortunately, these measurements did not
permit to check the validity of the standard models. High
sensitivity observations of these two SNRs in the soft
\gammaray~domain would then bring either a detection or strong
constraints on nucleosynthesis models in standard type Ia SNe. In
the case of Tycho, \cite{c:badenes03} have considered the
possibility to efficiently constrain the nature of the progenitor
through a careful analysis of the SNR thermal \xray~spectrum. They
have estimated a \ti{Ti} yield of about 6 $\times$ 10$^{-6}$
\msun. If one assumes such value for both Tycho and Kepler SNRs,
this would imply a \ti{Ti} line flux of $\sim$ 5 and 2 $\times$
10$^{-5}$ cm$^{-2}$ s$^{-1}$, respectively. Since these are
roughly the same as those expected from SN~1987A, a deep
observation with \simbolx~($\sim$ 10$^{6}$ s) would be
required\footnote{Note that, unlike SN~1987A, the non-negligible
size of Tycho and Kepler SNRs (8 and 3\m, respectively) may
require a few pointings to map them and an increase of the observing time.}.

\subsection{Search for Galactic "young, missing and hidden" SNRs: \ti{Ti} production and SNe rate}

The last topic that \simbolx~would be in a position to study
concerns the search for young and previously unknown Galactic
SNRs. Since four centuries and Kepler's SN, no stellar explosion
has been optically observed, while 2-3 such events are expected per century
 in the Milky Way \citep[see \eg][]{c:cappellaro99}.
The reason is that most of the SNe occurred in highly obscured
regions and therefore remain undetected. One way to avoid such a
limitation is to search for \ti{Ti} \gammaray~lines, exempt from
interstellar extinction. Moreover, 2-3 Cas~A/SN~1987A-like \ti{Ti}
events per century are required to explain the current Galactic
\ti{Ti} production rate deduced from the observed solar \ti{Ca}
abundance through the common Galactic chemical evolution models
\citep[see][and references therein]{c:the06}. However, several
\gammaray~line surveys
\citep{c:mahoney92,c:leising94,c:dupraz97,c:renaud04,c:the06} have
highlighted the problem of the lack of solid \ti{Ti}-source
candidates / young SNRs (except Cas~A), those that should have
occurred since Cas~A and are still not detected through the line
emission from \ti{Ti} decay. \cite{c:the06} have then raised the
question of the exceptionality of \ti{Ti}-producing SNe.

\begin{figure}[ht!]
\resizebox{\hsize}{!}{\includegraphics[clip=true]{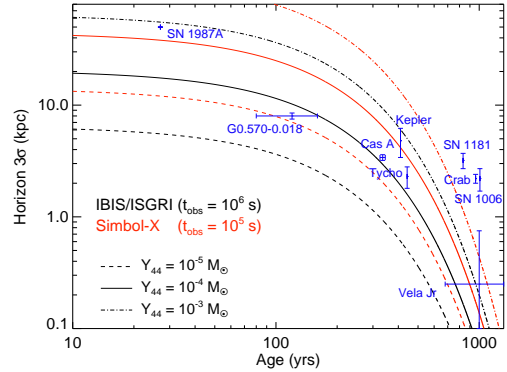}}

\caption{\footnotesize The 3 $\sigma$ sensitivity, expressed as
the distance at which a SNR would be detected at this significance
level (called horizon, in units of kpc), of \ibisgri~(black lines,
for an exposure time of 10$^{6}$ s) and \simbolx~(red lines, for
an exposure time of 10$^{5}$ s) as a function of the SNR age, for
\ti{Ti} yields of 10$^{-5}$ (dashed), 10$^{-4}$ (solid) and
10$^{-3}$ \msun~(dot-dashed). The historical SNRs are also
indicated with their respective uncertainties on the distance and
age \citep{c:green05}. Note that any SNR is considered as a
point-like source for both instruments in this calculation (which
is not the case for historical Galactic SNRs at the scale of the
\simbolx~PSF). G0.570-0.018 is a young SNR-candidate suggested by
\cite{c:senda02}.}

\label{f:fig4}

\end{figure}

Even if no serious candidate has emerged yet, significant
detections are expected with the continuous improvement of the
\ibisgri~survey sensitivity before the launch of \simbolx. The
relatively narrow field of view of \simbolx~will not permit a survey of the
Galaxy in the \ti{Ti} \gammaray~lines as that performed with
\integ. On the other hand, dedicated observations will allow a
gigantic step forward in sensitivity. Figure \ref{f:fig4}
illustrates this purpose and shows the 3 $\sigma$ sensitivity,
expressed as the distance at which a SNR would be detected at this
significance level, of \ibisgri~(for an exposure time of 10$^{6}$
s) and \simbolx~(for an exposure time of 10$^{5}$ s) as a function
of the SNR age for different \ti{Ti} yields. As an example,
G0.570-0.018 is a SNR-candidate close to the Galactic Center that
might be as young as $\sim$ 100 yr \citep{c:senda02}. Even if
\ibisgri~upper limits on the \ti{Ti} line flux and the
non-detection of radio emission brought severe constraints on the
nature of this source \citep{c:renaud06a}, \simbolx~should be able
to detect it for \ti{Ti} yields down to $\sim$ 10$^{-5}$ \msun.
Therefore, confirming these young Galactic SNR-candidates,
measuring precisely \ti{Sc} line fluxes and their
Doppler-velocity, and mapping the emission turn out to be
tasks that only \simbolx~can tackle.

\section{Conclusion}

\simbolx~will undoubtedly play a major role in our understanding
of the explosive nucleosynthesis processes through the \ti{Ti}
radioactive \gammaray~lines from young Galactic SNRs. Thanks to
its spectro-imaging capabilities and sensitivity, it will allow us
to locate precisely the \ti{Ti}-emitting regions and to measure
their velocity in Cas~A for the first time. The detection of these
\gammaray~lines in SN~1987A will greatly constraint the models of
evolution of massive stars and nucleosynthesis processes.
Moreover, it might detect them in other SNRs such as Tycho and
Kepler, whose progenitors are still a matter of debate. The issue
of the sources at the origin of the Galactic \ti{Ca} will also be
adressed by \simbolx~through dedicated observations of young
SNR-candidates that wide-field instruments like those onboard
\integ~should reveal.

\begin{acknowledgements}
We are grateful to the organisers of this nice \simbolx~workshop
in Bologna. M.R. would like to thank Philippe Ferrando and
Giuseppe Malaguti for their invitation.
\end{acknowledgements}

\bibliographystyle{aa}

\begin{thebibliography}{}

\bibitem[{{Ahmad} {et~al.}(2006){Ahmad}, {Greene}, {Moore}, {Ghelberg}, {Ofan},
  {Paul}, \& {Kutschera}}]{c:ahmad06}
{Ahmad}, I., {Greene}, J.~P., {Moore}, E.~F., {et~al.} 2006, \prc,
74, 065803

\bibitem[{{Anders} \& {Grevesse}(1989)}]{c:anders89}
{Anders}, E. \& {Grevesse}, N. 1989, \gca, 53, 197

\bibitem[{{Ashworth}(1980)}]{c:ashworth80}
{Ashworth}, Jr., W.~B. 1980, Journal for the History of Astronomy,
11, 1

\bibitem[{{Baade} \& {Minkowski}(1954)}]{c:baade54}
{Baade}, W. \& {Minkowski}, R. 1954, \apj, 119, 206

\bibitem[{{Badenes} {et~al.}(2003){Badenes}, {Bravo}, {Borkowski}, \&
  {Dom{\'{\i}}nguez}}]{c:badenes03}
{Badenes}, C., {Bravo}, E., {Borkowski}, K.~J., \&
{Dom{\'{\i}}nguez}, I. 2003,
  \apj, 593, 358

\bibitem[{{Bandiera}(1987)}]{c:bandiera87}
{Bandiera}, R. 1987, \apj, 319, 885

\bibitem[{{Cappellaro} {et~al.}(1999){Cappellaro}, {Evans}, \&
  {Turatto}}]{c:cappellaro99}
{Cappellaro}, E., {Evans}, R., \& {Turatto}, M. 1999, \aap, 351,
459

\bibitem[{{Decourchelle} {et~al.}(2001){Decourchelle}, {Sauvageot}, {Audard},
  {Aschenbach}, {Sembay}, {Rothenflug}, {Ballet}, {Stadlbauer}, \&
  {West}}]{c:decourchelle01}
{Decourchelle}, A., {Sauvageot}, J.~L., {Audard}, M., {et~al.}
2001, \aap, 365,
  L218

\bibitem[{{Dennefeld}(1982)}]{c:dennefeld82}
{Dennefeld}, M. 1982, \aap, 112, 215

\bibitem[{{Diehl} {et~al.}(2006){Diehl}, {Prantzos}, \& {von
  Ballmoos}}]{c:diehl06bis}
{Diehl}, R., {Prantzos}, N., \& {von Ballmoos}, P. 2006, Nuclear
Physics A,
  777, 70

\bibitem[{{Dupraz} {et~al.}(1997){Dupraz}, {Bloemen}, {Bennett}, {Diehl},
  {Hermsen}, {Iyudin}, {Ryan}, \& {Schoenfelder}}]{c:dupraz97}
{Dupraz}, C., {Bloemen}, H., {Bennett}, K., {et~al.} 1997, \aap,
324, 683

\bibitem[{{Fesen} {et~al.}(2006{\natexlab{a}}){Fesen}, {Hammell}, {Morse},
  {Chevalier}, {Borkowski}, {Dopita}, {Gerardy}, {Lawrence}, {Raymond}, \& {van
  den Bergh}}]{c:fesen06bbis}
{Fesen}, R.~A., {Hammell}, M.~C., {Morse}, J., {et~al.}
2006{\natexlab{a}},
  \apj, 636, 859

\bibitem[{{Fesen} {et~al.}(2006{\natexlab{b}}){Fesen}, {Hammell}, {Morse},
  {Chevalier}, {Borkowski}, {Dopita}, {Gerardy}, {Lawrence}, {Raymond}, \& {van
  den Bergh}}]{c:fesen06abis}
{Fesen}, R.~A., {Hammell}, M.~C., {Morse}, J., {et~al.}
2006{\natexlab{b}},
  \apj, 645, 283

\bibitem[{{Fransson} \& {Kozma}(2002)}]{c:fransson02}
{Fransson}, C. \& {Kozma}, C. 2002, New Astronomy Review, 46, 487

\bibitem[{{Gaensler} {et~al.}(2007){Gaensler}, {Staveley-Smith}, {Manchester},
  {Kesteven}, {Ball}, \& {Tzioumis}}]{c:gaensler07}
{Gaensler}, B.~M., {Staveley-Smith}, L., {Manchester}, R.~N.,
{et~al.} 2007,
  ArXiv e-prints, 705

\bibitem[{{Green}(2005)}]{c:green05}
{Green}, D.~A. 2005, Memorie della Societa Astronomica Italiana,
76, 534

\bibitem[{{Horoi} {et~al.}(2002){Horoi}, {Jora}, {Zelevinsky}, {Murphy},
  {Boyd}, \& {Rauscher}}]{c:horoi02}
{Horoi}, M., {Jora}, R., {Zelevinsky}, V., {et~al.} 2002, \prc,
66, 015801

\bibitem[{{Hwang} {et~al.}(2002){Hwang}, {Decourchelle}, {Holt}, \&
  {Petre}}]{c:hwang02}
{Hwang}, U., {Decourchelle}, A., {Holt}, S.~S., \& {Petre}, R.
2002, \apj, 581,
  1101

\bibitem[{{Hwang} \& {Laming}(2003)}]{c:hwang03}
{Hwang}, U. \& {Laming}, J.~M. 2003, \apj, 597, 362

\bibitem[{{Hwang} {et~al.}(2004){Hwang}, {Laming}, {Badenes}, {Berendse},
  {Blondin}, {Cioffi}, {DeLaney}, {Dewey}, {Fesen}, {Flanagan}, {Fryer},
  {Ghavamian}, {Hughes}, {Morse}, {Plucinsky}, {Petre}, {Pohl}, {Rudnick},
  {Sankrit}, {Slane}, {Smith}, {Vink}, \& {Warren}}]{c:hwang04}
{Hwang}, U., {Laming}, J.~M., {Badenes}, C., {et~al.} 2004, \apjl,
615, L117

\bibitem[{{Iwamoto} {et~al.}(1999){Iwamoto}, {Brachwitz}, {Nomoto},
  {Kishimoto}, {Umeda}, {Hix}, \& {Thielemann}}]{c:iwamoto99}
{Iwamoto}, K., {Brachwitz}, F., {Nomoto}, K., {et~al.} 1999,
\apjs, 125, 439

\bibitem[{{Iyudin} {et~al.}(1994){Iyudin}, {Diehl}, {Bloemen}, {Hermsen},
  {Lichti}, {Morris}, {Ryan}, {Schoenfelder}, {Steinle}, {Varendorff}, {de
  Vries}, \& {Winkler}}]{c:iyudin94}
{Iyudin}, A.~F., {Diehl}, R., {Bloemen}, H., {et~al.} 1994, \aap,
284, L1

\bibitem[{{Iyudin} {et~al.}(1998){Iyudin}, {Schonfelder}, {Bennett}, {Bloemen},
  {Diehl}, {Hermsen}, {Lichti}, {van der Meulen}, {Ryan}, \&
  {Winkler}}]{c:iyudin98}
{Iyudin}, A.~F., {Schonfelder}, V., {Bennett}, K., {et~al.} 1998,
\nat, 396,
  142

\bibitem[{{Kinugasa} \& {Tsunemi}(1994)}]{c:kinugasa94}
{Kinugasa}, K. \& {Tsunemi}, H. 1994, in New Horizon of X-Ray
Astronomy. First
  Results from ASCA, ed. F.~{Makino} \& T.~{Ohashi}, 469--+

\bibitem[{{Kurfess} {et~al.}(1992){Kurfess}, {Johnson}, {Kinzer}, {Kroeger},
  {Strickman}, {Grove}, {Leising}, {Clayton}, {Grabelsky}, {Purcell}, {Ulmer},
  {Cameron}, \& {Jung}}]{c:kurfess92}
{Kurfess}, J.~D., {Johnson}, W.~N., {Kinzer}, R.~L., {et~al.}
1992, \apjl, 399,
  L137

\bibitem[{{Laming} \& {Hwang}(2003)}]{c:laming03}
{Laming}, J.~M. \& {Hwang}, U. 2003, \apj, 597, 347

\bibitem[{{Leising} \& {Share}(1994)}]{c:leising94}
{Leising}, M.~D. \& {Share}, G.~H. 1994, \apj, 424, 200

\bibitem[{{Limongi} \& {Chieffi}(2003)}]{c:lc03}
{Limongi}, M. \& {Chieffi}, A. 2003, \apj, 592, 404

\bibitem[{{Livne} \& {Arnett}(1995)}]{c:livne95}
{Livne}, E. \& {Arnett}, D. 1995, \apj, 452, 62

\bibitem[{{Lodders}(2003)}]{c:lodders03}
{Lodders}, K. 2003, \apj, 591, 1220

\bibitem[{{Maeda}(2006)}]{c:maeda06}
{Maeda}, K. 2006, \apj, 644, 385

\bibitem[{{Mahoney} {et~al.}(1992){Mahoney}, {Ling}, {Wheaton}, \&
  {Higdon}}]{c:mahoney92}
{Mahoney}, W.~A., {Ling}, J.~C., {Wheaton}, W.~A., \& {Higdon},
J.~C. 1992,
  \apj, 387, 314

\bibitem[{{Matz} {et~al.}(1988){Matz}, {Share}, {Leising}, {Chupp}, \&
  {Vestrand}}]{c:matz88}
{Matz}, S.~M., {Share}, G.~H., {Leising}, M.~D., {Chupp}, E.~L.,
\& {Vestrand},
  W.~T. 1988, \nat, 331, 416

\bibitem[{{Morris} \& {Podsiadlowski}(2007)}]{c:morris07}
{Morris}, T. \& {Podsiadlowski}, P. 2007, Science, 315, 1103

\bibitem[{{Motizuki} \& {Kumagai}(2004)}]{c:mochizuki04}
{Motizuki}, Y. \& {Kumagai}, S. 2004, New Astronomy Review, 48, 69

\bibitem[{{Nassar} {et~al.}(2006){Nassar}, {Paul}, {Ahmad}, {Ben-Dov},
  {Caggiano}, {Ghelberg}, {Goriely}, {Greene}, {Hass}, {Heger}, {Heinz},
  {Henderson}, {Janssens}, {Jiang}, {Kashiv}, {Nara Singh}, {Ofan}, {Pardo},
  {Pennington}, {Rehm}, {Savard}, {Scott}, \& {Vondrasek}}]{c:nassar06}
{Nassar}, H., {Paul}, M., {Ahmad}, I., {et~al.} 2006, Physical
Review Letters,
  96, 041102

\bibitem[{{Podsiadlowski}(1992)}]{c:podsiadlowski92}
{Podsiadlowski}, P. 1992, \pasp, 104, 717

\bibitem[{{Rauscher} {et~al.}(2002){Rauscher}, {Heger}, {Hoffman}, \&
  {Woosley}}]{c:rhhw02}
{Rauscher}, T., {Heger}, A., {Hoffman}, R.~D., \& {Woosley}, S.~E.
2002, \apj,
  576, 323

\bibitem[{{Reed} {et~al.}(1995){Reed}, {Hester}, {Fabian}, \&
  {Winkler}}]{c:reed95}
{Reed}, J.~E., {Hester}, J.~J., {Fabian}, A.~C., \& {Winkler},
P.~F. 1995,
  \apj, 440, 706

\bibitem[{{Renaud} {et~al.}(2004){Renaud}, {Lebrun}, {Ballet}, {Decourchelle},
  {Terrier}, \& {Prantzos}}]{c:renaud04}
{Renaud}, M., {Lebrun}, F., {Ballet}, J., {et~al.} 2004, in ESA
SP-552: 5th
  INTEGRAL Workshop on the INTEGRAL Universe, ed. V.~{Schoenfelder},
  G.~{Lichti}, \& C.~{Winkler}, 81--+

\bibitem[{{Renaud} {et~al.}(2006{\natexlab{a}}){Renaud}, {Paron}, {Terrier},
  {Lebrun}, {Dubner}, {Giacani}, \& {Bykov}}]{c:renaud06a}
{Renaud}, M., {Paron}, S., {Terrier}, R., {et~al.}
2006{\natexlab{a}}, \apj

\bibitem[{{Renaud} {et~al.}(2006{\natexlab{b}}){Renaud}, {Vink},
  {Decourchelle}, {Lebrun}, {Hartog}, {Terrier}, {Couvreur}, {Kn{\"o}dlseder},
  {Martin}, {Prantzos}, {Bykov}, \& {Bloemen}}]{c:renaud06e}
{Renaud}, M., {Vink}, J., {Decourchelle}, A., {et~al.}
2006{\natexlab{b}},
  \apjl, 647, L41

\bibitem[{{Renaud} {et~al.}(2006{\natexlab{c}}){Renaud}, {Vink},
  {Decourchelle}, {Lebrun}, {Terrier}, \& {Ballet}}]{c:renaud06b}
{Renaud}, M., {Vink}, J., {Decourchelle}, A., {et~al.}
2006{\natexlab{c}}, New
  Astronomy Review

\bibitem[{{Reynolds} {et~al.}(2007){Reynolds}, {Borkowski}, {Hwang}, {Hughes},
  {Badenes}, {Laming}, \& {Blondin}}]{c:reynolds07}
{Reynolds}, S.~P., {Borkowski}, K.~J., {Hwang}, U., {et~al.} 2007,
ArXiv
  e-prints, 708

\bibitem[{{Reynoso} {et~al.}(1999){Reynoso}, {Vel{\'a}zquez}, {Dubner}, \&
  {Goss}}]{c:reynoso99}
{Reynoso}, E.~M., {Vel{\'a}zquez}, P.~F., {Dubner}, G.~M., \&
{Goss}, W.~M.
  1999, \aj, 117, 1827

\bibitem[{{R{\"o}pke} {et~al.}(2005){R{\"o}pke}, {Gieseler}, \&
  {Hillebrandt}}]{c:ropke05}
{R{\"o}pke}, F.~K., {Gieseler}, M., \& {Hillebrandt}, W. 2005, in
ASP Conf.
  Ser. 342: 1604-2004: Supernovae as Cosmological Lighthouses, ed.
  M.~{Turatto}, S.~{Benetti}, L.~{Zampieri}, \& W.~{Shea}, 397--+

\bibitem[{{Sch{\"o}nfelder} {et~al.}(2000){Sch{\"o}nfelder}, {Bloemen},
  {Collmar}, {Diehl}, {Hermsen}, {Kn{\"o}dlseder}, {Lichti}, {Pl{\"u}schke},
  {Ryan}, {Strong}, \& {Winkler}}]{c:schonfelder00}
{Sch{\"o}nfelder}, V., {Bloemen}, H., {Collmar}, W., {et~al.}
2000, in American
  Institute of Physics Conference Series, ed. M.~L. {McConnell} \& J.~M.
  {Ryan}, 54--+

\bibitem[{{Senda} {et~al.}(2002){Senda}, {Murakami}, \& {Koyama}}]{c:senda02}
{Senda}, A., {Murakami}, H., \& {Koyama}, K. 2002, \apj, 565, 1017

\bibitem[{{Smith}(2007)}]{c:smith07}
{Smith}, N. 2007, \aj, 133, 1034

\bibitem[{{Smith} {et~al.}(1991){Smith}, {Kirshner}, {Blair}, \&
  {Winkler}}]{c:smith91}
{Smith}, R.~C., {Kirshner}, R.~P., {Blair}, W.~P., \& {Winkler},
P.~F. 1991,
  \apj, 375, 652

\bibitem[{{Sonzogni} {et~al.}(2000){Sonzogni}, {Rehm}, {Ahmad}, {Borasi},
  {Bowers}, {Brumwell}, {Caggiano}, {Davids}, {Greene}, {Harss}, {Heinz},
  {Henderson}, {Janssens}, {Jiang}, {McMichael}, {Nolen}, {Pardo}, {Paul},
  {Schiffer}, {Segel}, {Seweryniak}, {Siemssen}, {Truran}, {Uusitalo},
  {Wiedenh{\"o}ver}, \& {Zabransky}}]{c:sonzogni00}
{Sonzogni}, A.~A., {Rehm}, K.~E., {Ahmad}, I., {et~al.} 2000,
Physical Review
  Letters, 84, 1651

\bibitem[{{Stephenson} \& {Green}(2002)}]{c:stephenson02}
{Stephenson}, F.~R. \& {Green}, D.~A. 2002, {Historical supernovae
and their
  remnants, by F. Richard Stephenson and David A. Green. International series
  in astronomy and astrophysics, vol. 5. Oxford: Clarendon Press, 2002, ISBN
  0198507666}

\bibitem[{{The} {et~al.}(2006){The}, {Clayton}, {Diehl}, {Hartmann}, {Iyudin},
  {Leising}, {Meyer}, {Motizuki}, \& {Sch{\"o}nfelder}}]{c:the06}
{The}, L.-S., {Clayton}, D.~D., {Diehl}, R., {et~al.} 2006, \aap,
450, 1037

\bibitem[{{The} {et~al.}(1998){The}, {Clayton}, {Jin}, \& {Meyer}}]{c:the98}
{The}, L.-S., {Clayton}, D.~D., {Jin}, L., \& {Meyer}, B.~S. 1998,
\apj, 504,
  500

\bibitem[{{Thielemann} {et~al.}(1996){Thielemann}, {Nomoto}, \&
  {Hashimoto}}]{c:tnh96}
{Thielemann}, F.-K., {Nomoto}, K., \& {Hashimoto}, M.-A. 1996,
\apj, 460, 408

\bibitem[{{Thorstensen} {et~al.}(2001){Thorstensen}, {Fesen}, \& {van den
  Bergh}}]{c:thorstensen01}
{Thorstensen}, J.~R., {Fesen}, R.~A., \& {van den Bergh}, S. 2001,
\aj, 122,
  297

\bibitem[{{Travaglio} {et~al.}(2004){Travaglio}, {Hillebrandt}, {Reinecke}, \&
  {Thielemann}}]{c:travaglio04}
{Travaglio}, C., {Hillebrandt}, W., {Reinecke}, M., \&
{Thielemann}, F.-K.
  2004, \aap, 425, 1029

\bibitem[{{Tueller} {et~al.}(1990){Tueller}, {Barthelmy}, {Gehrels},
  {Teegarden}, {Leventhal}, \& {MacCallum}}]{c:tueller90}
{Tueller}, J., {Barthelmy}, S., {Gehrels}, N., {et~al.} 1990,
\apjl, 351, L41

\bibitem[{{Vink}(2004)}]{c:vink04}
{Vink}, J. 2004, New Astronomy Review, 48, 61

\bibitem[{{Vink} {et~al.}(1998){Vink}, {Bloemen}, {Kaastra}, \&
  {Bleeker}}]{c:vink98}
{Vink}, J., {Bloemen}, H., {Kaastra}, J.~S., \& {Bleeker},
J.~A.~M. 1998, \aap,
  339, 201

\bibitem[{{Vink} {et~al.}(2001){Vink}, {Laming}, {Kaastra}, {Bleeker},
  {Bloemen}, \& {Oberlack}}]{c:vink01}
{Vink}, J., {Laming}, J.~M., {Kaastra}, J.~S., {et~al.} 2001,
\apjl, 560, L79

\bibitem[{{Woosley} {et~al.}(1973){Woosley}, {Arnett}, \&
  {Clayton}}]{c:woosley73}
{Woosley}, S.~E., {Arnett}, W.~D., \& {Clayton}, D.~D. 1973,
\apjs, 26, 231

\bibitem[{{Woosley} {et~al.}(1995){Woosley}, {Langer}, \& {Weaver}}]{c:wlw95}
{Woosley}, S.~E., {Langer}, N., \& {Weaver}, T.~A. 1995, \apj,
448, 315

\bibitem[{{Woosley} \& {Weaver}(1994)}]{c:woosley94}
{Woosley}, S.~E. \& {Weaver}, T.~A. 1994, \apj, 423, 371

\bibitem[{{Woosley} \& {Weaver}(1995)}]{c:ww95}
{Woosley}, S.~E. \& {Weaver}, T.~A. 1995, \apjs, 101, 181

\bibitem[{{Young} {et~al.}(2006){Young}, {Fryer}, {Hungerford}, {Arnett},
  {Rockefeller}, {Timmes}, {Voit}, {Meakin}, \& {Eriksen}}]{c:young06}
{Young}, P.~A., {Fryer}, C.~L., {Hungerford}, A., {et~al.} 2006,
\apj, 640, 891

\end{thebibliography}

\end{document}